\documentclass[11pt]{article}
\usepackage{amssymb}
\usepackage{amsmath}
\usepackage[dvips]{epsfig}
\begin{document}

\title{Eigenmodes of index-modulated layers with lateral PMLs}
\author{D. C. Skigin \footnote{Member of Conicet}\\
{\em Grupo de Electromagnetismo Aplicado,}\\
{\em Departamento de F\'{\i}sica,}\\
{\em Facultad de Ciencias Exactas y Naturales, }\\
{\em Universidad de Buenos Aires, }\\
{\em Ciudad Universitaria, Pabell\'{o}n I, }\\
{\em C1428EHA Buenos Aires, Argentina}}
\date{}
\maketitle

\baselineskip 4.5ex

\section*{Abstract}

Maxwell equations are solved in a layer comprising a finite number of
homogeneous isotropic dielectric regions ended by anisotropic perfectly
matched layers (PMLs). 
The boundary-value problem is solved and the dispersion relation inside the PML is derived. 
The general expression of the eigenvalues equation for an arbitrary number of regions in each 
layer is obtained, and both polarization modes are considered.
The modal functions of a single layer ended by PMLs are found, and their
orthogonality relation is derived. 
The present method is useful to simulate scattering problems from dielectric objects 
as well as propagation in planar slab waveguides. Its potential to deal with
more complex problems such as the scattering from an object with arbitrary cross 
section in open space using the multilayer modal method is briefly discussed. 

\newpage

\section{Introduction}

The perfectly matched layer (PML) is a fictitious material which does not
reflect incident propagating waves regardless of the incident angle,
frequency and polarization. It was firstly introduced by Berenger \cite
{berenger1,berenger2} as a useful absorbing boundary condition to truncate
the computational domain in finite-difference time-domain applications. A
different formulation of the PML, given by Sacks et al. \cite{Sacks}, is
based on exploiting constitutive characteristics of anisotropic materials to
provide a reflectionless interface. This formulation
offers the special advantage that it does not require modification of
Maxwell equations \cite{Sacks}. Both formulations of the PML are very
popular among the electrical engineering community and their use in the
optics community has been growing in the last few years 
\cite{PML_JOSAB}-\cite{PML2}. PMLs look attractive, mainly because of 
their potential capacity to simulate open-structure problems by bounded
computational domains, which consequently reduce significantly the computational
costs involved in the calculations \cite{Rogier1}-\cite{Rogier3}. Even 
though this kind of medium is
usually used in the framework of finite-element methods \cite{Gedney}-\cite{Tischler}, 
there are also studies which incorporate the anisotropic absorber in different 
approaches \cite{Merle2}, \cite{Derudder0}-\cite{Olyslager1}. In particular, 
the modal approach appears as an interesting alternative for the description
of the fields in index-modulated structures since it highlights the physics
of the problem.

The modal approach has been applied by many authors to dielectric lamellar gratings
in classical \cite{Knop}-\cite{Botten0} as well as in conical mounting \cite{Peng1,Li}.
In his work, Li \cite{Li} derived rigorously the eigenfunctions and also their completeness
and ortogonality relations from the boundary-value problem. Later on, this formalism was
extended to deal with multiply grooved lamellar gratings \cite{Miller}, where the eigenvalues
equation for an arbitrary number of different dielectric zones was obtained and carefully
analyzed for real and complex refraction indices. The modal formulation presented in \cite{Miller}
was then extended to deal with finite structures such as index-modulated apertures \cite{Kuittinen}.

A further step on the development of the modal method was the application of the multilayer
approximation \cite{Peng2} not only to infinite gratings \cite{Li2} but also to finite
structures with arbitrary shapes of the corrugations \cite{diana9}-\cite{diana12}. 
However, all these works were restricted to structures laterally closed by perfect conductors.
If illuminated by finite beams, these structures can approximately simulate the
scattering from purely dielectric finite structures \cite{Kuittinen}, \cite{diana9}-\cite{diana12}.

\begin{figure}[h]
\begin{center}
\includegraphics[width=7in]{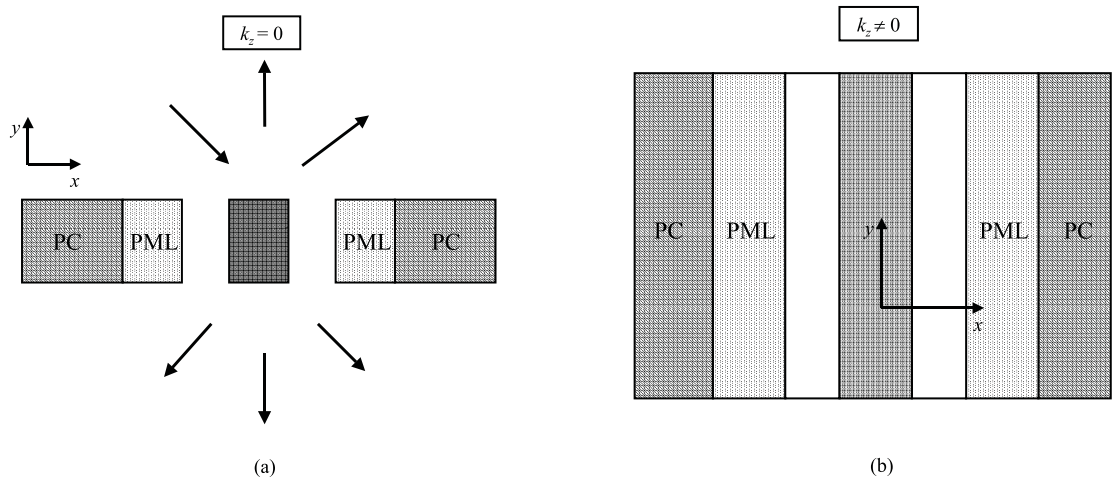}
\end{center}
\caption{(a) Scheme of the scattering problem by an index-modulated aperture; (b) 
Scheme of the propagation problem in a planar slab waveguide.}
\end{figure}

It is well known that either the pseudeperiodic condition in infinite gratings or 
the assumption of perfect conductivity in finite structures forces the eigenvalues set
to be discrete. On the other hand, open problems like propagation in dielectric
waveguides or scattering from dielectric objects in open space, have a continuum set of
eigenvalues. Therefore, it is allways convenient to limit the problem domain, in order to
avoid large volumes of calculus. This can be done by means of perfectly matched layers,
which ensure absorption and attenuation of the incident radiation and consequently
simulate the open space better. A detailed formulation of the discretization of the
continuous spectrum by means of a Dirichlet boundary condition for the complex plane
has been recently presented \cite{Olyslager2}, where the solution of the 
Sturm-Liouville problem for a layered structure is based on the PML boundary condition. 
The completeness of the eigenmodes of parallel plate waveguides with PML terminations 
has also been proven \cite{Knockaert}.

The purpose of this paper is to find the analytical expressions of the eigenmodes of a single index-modulated layer ended by PMLs at both sides, and also to derive the orthogonality 
relations that satisfy 
the set of eigenfunctions. A modal formulation is presented to apply
the eigenmodes expressions to the solution of the scattering problem
by a two-dimensional object of rectangular cross section in open space. 
We cover simultaneously the eigenmodes of an index-modulated 
aperture (schematized in Fig. 1a) as well as those of a planar slab 
waveguide (Fig. 1b), since both problems have an equivalent formulation.
Numerical examples comparing the resulting eigenvalues with those
found by a different approach \cite{Derudder2} are shown. The present paper
is more general than Ref. \cite{Derudder2} since we find the eigenvalues and 
the eigenmodes of a layer comprising an arbitrary number of different homogeneous 
regions, with the aim of extending the modal method to more general structures ended by PMLs.

Finally, a brief discussion on the potential applicability of the modal method
to the scattering problem by a 2-D object of arbitrary cross section in open space is given.
This extension can be made by means of the multilayer approximation \cite{Peng2}, and 
the application of any propagation algorithm such as the R-matrix 
\cite{Merle1, Merle2, Li2}\cite{Li3}-\cite{Neviere}.

\section{Modal formulation}

In this section we develop a rigorous method to find the exact modes of a structure 
as that of Fig. 2.
The modes of such a structure can be then used to find the solution of waveguide problems and
scattering problems.  

\begin{figure}[h]
\begin{center}
\includegraphics[width=5in]{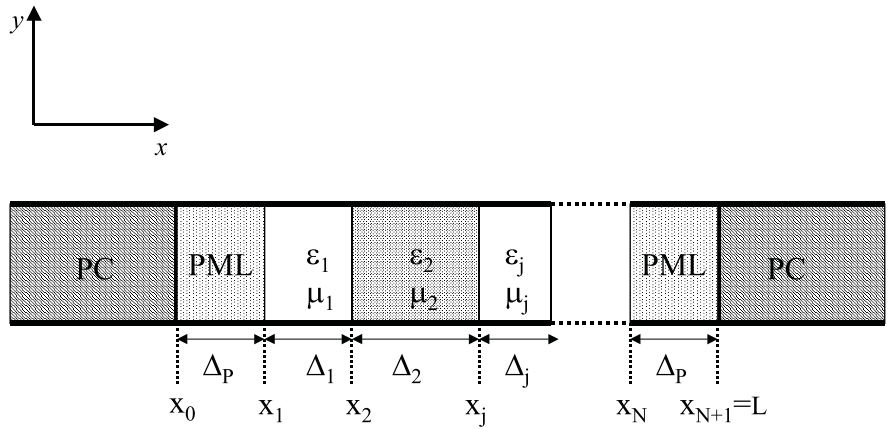}
\end{center}
\caption{Configuration of the problem.}
\end{figure}

Consider an index-modulated layer invariant in the $\hat{z}$ direction, as shown in Fig. 2. 
Each homogeneous zone 
$x_j \leq x \leq x_j + \Delta_j$ of width $\Delta_j$ has a permittivity $\epsilon_j$ 
and a permeability $\mu_j$, where $\epsilon_j$ and $\mu_j$ can be complex numbers. The layer is 
ended at both sides by perfectly matched
layers backed by perfect conductors. The PML is a particular anisotropic medium 
especially designed to absorb the received radiation \cite{Sacks}, and its constitutive parameters 
are $\tilde \epsilon_{{\rm PML}}=\epsilon \tilde \Lambda$ and  
$\tilde \mu_{{\rm PML}}=\mu \tilde \Lambda$ where $\epsilon$ and $\mu$ 
are the permittivity and permeability of the adjacent isotropic zone, respectively.
For an interface parallel to the $(y,z)$-plane, the tensor $\tilde \Lambda$ must be defined as: 
\begin{equation}
\tilde \Lambda= \left[ 
\begin{array}{ccc}
1/b & 0 & 0 \\ 
0 & b & 0 \\ 
0 & 0 & b
\end{array}
\right]\;\;.
\end{equation}
The parameter $b$ is a complex number: its real part is related to the wavelength inside the PML, and the
imaginary part accounts for the losses in the material, i.e., the
attenuation of the propagating waves.

As stated above, to find the eigenmodes of the structure depicted in Fig. 2
can be useful for waveguide and scattering problems. We are interested in studying the propagation of 
plane waves along the structure, and then a time dependence $e^{-i\omega t}$ ($\omega$ being the
frequency of the plane wave) is implied and suppressed in all the paper. Since the structure is
invariant in the $z$-direction, the problem can be separated into two scalar problems for
$E_z$ (TM polarization) and $H_z$ (TE polarization). 
Then, the $x$- and $y$- components of the fields can be written
in terms of the $z$-components \cite{Jackson}.

Within the PML, Maxwell curl equations are
\begin{equation}
\vec{\nabla} \times \vec{E} = i \,\frac{\omega}{c}\, \mu \, \tilde \Lambda \,\vec{H}\;\;,   \label{maxwell1}
\end{equation}
\begin{equation}
\vec{\nabla} \times \vec{H} = -i\, \frac{\omega}{c} \,  \epsilon \, \tilde \Lambda \, \vec{E}\;\;,   \label{maxwell2}
\end{equation}
where $c$ is the speed of light in vacuum and $\vec{E}$ and $\vec{H}$ are the electric and magnetic
fields, respectively. In the case of a slab waveguide (Fig. 1b), we are interested in finding the solutions 
of Maxwell equations that have a $z$-dependence of the form $e^{i k_z z}$. Then, 
combining eqs. (\ref{maxwell1}) and (\ref{maxwell2}), and taking into 
account the invariance of the problem along the $\hat{z}$ direction, we get the propagation 
equations for a layer with $x$-dependent constitutive parameters, for both polarization modes:
\begin{equation}
\frac{\mu(x)}{b(x)} \frac{\partial}{\partial x}\left[ \frac{1}{b(x) \mu(x)} \frac{\partial E_z(x,y)}{\partial x} 
\right]+ \frac{\partial^2 E_z(x,y)}{\partial y^2} +
\gamma^2(x) E_z(x,y)  =0\;\;,						\label{helmholtzE}
\end{equation}
\begin{equation}
\frac{\epsilon(x)}{b(x)} \frac{\partial}{\partial x}\left[ \frac{1}{b(x) \epsilon(x)} \frac{\partial H_z(x,y)}{\partial x} \right]
+ \frac{\partial^2 H_z(x,y)}{\partial y^2} +
\gamma^2(x) H_z(x,y)  =0\;\;,						\label{helmholtzH}
\end{equation}
where 
\begin{equation}
\gamma^2(x)=\frac{\omega^2}{c^2} \epsilon(x) \mu(x) - k_z^2\;\;,			\label{gamma}
\end{equation}
and $b(x)=1$ in the isotropic regions. For the scattering problem of a plane wave with wave vector $\vec{k}$ 
in the ($x,y$)-plane 
impinging on the structure (Fig. 1a), $k_z=0$ and then $\gamma^2(x)=k^2(x)=(\omega^2/c^2) \epsilon(x) \mu(x)$. 
On the other hand, for a slab waveguide the problem is invariant in the $\hat{y}$ direction, and then 
$\partial_y=0$. Since the form of eqs. (\ref{helmholtzE}) and (\ref{helmholtzH})
is identical, we unify the treatment of both polarizations in a single differential equation
\begin{equation}
\frac{\sigma(x)}{b(x)} \frac{\partial}{\partial x}\left[ \frac{1}{b(x) \sigma(x)} \frac{\partial \psi(x,y)}{\partial x} \right]
+ \frac{\partial^2 \psi(x,y)}{\partial y^2} +
\gamma^2(x) \psi(x,y)  =0\;\;,						\label{helmholtz-comun}
\end{equation} 
where 
\begin{equation}
\sigma(x)=\left\{\begin{array}{ll}
		\mu(x) & \mbox{for TM polarization}\\
\\
		\epsilon(x) & \mbox{for TE polarization}     \label{sigma}
\end{array}
\right.\;\;,
\end{equation}
and $\psi$ represents either $E_z$ (TM case) or $H_z$ (TE case).

The eigenmodes of the structure are the set of linearly independent functions that satisfy 
by themselves the boundary conditions at all the interfaces $x=x_j$, $j=0,...,N+1$, and form a 
complete basis.
In particular, since the layer is ended by a perfect conductor, we require that the tangential
component of the electric field vanishes at $x=x_0$ and at $x=L$.
To do so, we first solve eq. (\ref{helmholtz-comun}) in each region 
(isotropic and anisotropic) and then match these partial solutions correspondingly. For the 
most general case in which $\partial_y \neq 0$ a separated solution is proposed: 

\begin{equation}
\psi(x,y)=X(x).Y(y)\;\;,			\label{psi}
\end{equation}
and substituting eq. (\ref{psi}) in eq. (\ref{helmholtz-comun}) we get two ordinary differential equations
for the functions $X(x)$ and $Y(y)$
\begin{eqnarray}
\frac{d^2 Y}{dy^2} + v^2 Y&=&0\;\;,          \label{ec-y}\\
\nonumber \\						
\frac{\sigma(x)}{b(x)} \frac{d}{d x}\left[ \frac{1}{b(x) \sigma(x)} \frac{d X}{d x}\right] 
+ (\gamma^2(x) - v^2) X &=&0\;\;,		\label{ec-x}						
\end{eqnarray} 
where $v$ is a constant. The formal solution of eq. (\ref{ec-y}) is straightforward:
\begin{equation}
Y(y)=a \cos(v y) + b \sin(v y)\;\;,             \label{y}
\end{equation} 
and we will now focus on the differential equation for $X(x)$.
Notice that for the scattering problem $\gamma^2(x)=k^2(x)-v^2$ and for the slab waveguide problem 
$\gamma^2(x)=k^2(x)-k_z^2$. Then, we can unify both cases by setting $\rho=v^2$ in the first case and
$\rho=k_z^2$ in the second case.

Eq. (\ref{ec-x}) together with the boundary conditions at the 
ends of the layer ($x=x_0$ and $x=L$) pose a boundary-value problem
\begin{equation}
{\cal L} X = \rho X\;\;,			\label{sturm}
\end{equation}
where $\cal L$ is the differential operator 
\begin{equation}
{\cal L}=\frac{\sigma(x)}{b(x)} \frac{d}{d x}\left[ \frac{1}{b(x) \sigma(x)} \frac{d}{d x}\right]+ 
\frac{\omega^2}{c^2} \mu(x) \epsilon(x) \;\;.					\label{L}
\end{equation}
Since $b$ is a complex number, the operator $\cal L$ is a non-self-adjoint operator 
\cite{DiffEcs}. From the theory of non self-adjoint problems we know that the eigenvalues
are infinite and complex in general, and the eigenfunctions do not necessarily form a 
complete and orthogonal set \cite{DiffEcs}. To find a useful solution to our problem, is then necessary to consider the adjoint problem of (\ref{sturm})
\begin{equation}
{\cal L}^* X^\dag = \rho^\dag X^\dag\;\;,			\label{sturm-adjunto}
\end{equation} 
where the asterisk denotes complex conjugate and the $^\dag$ denotes adjoint. The eigenvalues of the
adjoint problem are the complex conjugates of those of the direct problem, i.e.,
\begin{equation}
\rho^\dag = \rho^*\;\;,
\end{equation}
and the sets of eigenfunctions \{$X_m(x)$\} and \{$X_n^\dag(x)$\} form a bi-orthonormal set such that $(X_m, X_n^\dag)=\delta_{mn}$, where the internal product $(X_m, X_n^\dag)$ is defined as
\begin{equation}
(X_m, X_n^\dag)=\int \frac{b(x)}{\sigma(x)} X_m(x) (X_n^\dag(x))^* dx\;\;.
\end{equation}
Then, any continuous and piecewise differentiable function that satisfies the boundary conditions
at the ends of the layer can be expanded as
\begin{equation}
f(x)=\sum_m A_m X_m(x)\;\;. 
\end{equation}
Taking into account these facts, we are going to find the solutions of eq. (\ref{ec-x}). 
In each homogeneous region ($j$), we propose a solution of the form 
\begin{equation}
X_j(x)=A_j \cos[u_j (x-x_j)] + B_j \sin[u_j (x-x_j)]\quad \mbox{for $x_j\,<\,x\,<\,x_{j}+\Delta_j$}\;\;, 			\label{sol-x}
\end{equation}
and substitute it into eq. (\ref{ec-x}) to get the dispersion relation 
\begin{equation}
\frac{u_j^2}{b_j^2}+\rho=\gamma_j^2\;\;,			\label{dispersion} 
\end{equation}
for the two kinds of zones we have in the structure: isotropic ($b_j=1$) and anisotropic 
($b_j \in C$, $b_j \neq 1$). Notice that $\rho$ is a constant for each eigenmode, and the 
constitutive parameters of the PML
satisfy $\tilde \epsilon_0=\epsilon_1 \tilde \Lambda$, 
$\tilde \mu_0=\mu_1 \tilde \Lambda$, and then $k_0^2=k_1^2$. 
The same occurs at the other end of the layer, and therefore,
$u_0 = b\, u_1$ and $u_N = b\, u_{N-1}$, i.e., the eigenvalue in the PML region is $b$ times
the eigenvalue in the adjacent isotropic region.
Then, the solution of eq. (\ref{ec-x}) in the whole layer can be expressed in terms of harmonic
functions,
where the amplitudes are such that satisfy the boundary conditions at the vertical interfaces, 
and the eigenvalues $u$ and $\rho$ satisfy the corresponding dispersion relation, depending on the
zone (eq. (\ref{dispersion})). Defining $\tilde{X}(x)$ as:
\begin{equation}
\tilde{X}(x) = \frac{u_{j}}{\sigma_j}\;[-B_{j} \cos{[u_{j} (x-x_{j})]}
+ A_{j} \sin{[u_{j} (x-x_{j})]}]\;\;\;  \mbox{for $x_{j}\,<\,x\,<\,x_{j}+\Delta_j$}
\;\;,
\end{equation}
it is easy to verify that \cite{Kuittinen}:
\begin{eqnarray}
X(x)&=&X(x_{j})\;\cos(u_{j} (x-x_{j})) -  
\frac{\sigma_j}{u_{j}}\tilde{X}(x_{j})\;\sin(u_{j} (x-x_{j}))\;,\;\;\;  \mbox{for $x_{j}\,<\,x\,<\,x_{j}+\Delta_j$} \nonumber\\
\tilde{X}(x)&=&\tilde{X}(x_{j})\;\cos(u_{j} (x-x_{j})) +
\frac{u_{j}}{\sigma_j} X(x_{j})\;\sin(u_{j} (x-x_{j}))\;,\;\;\;  \mbox{for $x_{j}\,<\,x\,<\,x_{j}+\Delta_j$}\;. \label{uu}
\end{eqnarray}
Notice that $\tilde{X}(x)$ is proportional to the normal derivative of
$X(x)$ at the vertical interfaces, which implies that it represents the
tangential component of the magnetic (electric) field in the case of TM (TE)
polarization.

Equations (\ref{uu}) provide us with a relation between two field
quantities at both sides of each homogeneous zone bounded by the interfaces at
$x=x_{j}$ and $x=x_{j}+\Delta_j$. This relation can be expressed in matrix form as:
\begin{equation}
\left(
\begin{array}{cc}
{\bf X}(x_{j+1})\\
\\ 
\tilde{\bf X}(x_{j+1})
\end{array}
\right) =
{\bf M}_{j}(\rho)
\left(
\begin{array}{cc}
{\bf X}(x_{j}) \\
\\ 
\tilde{\bf X}(x_{j})
\end{array}
\right)\;\;,  \label{relm}
\end{equation}
where ${\bf M}_{j}(\rho)$ is a matrix given by:
\begin{equation}
{\bf M}_{j}(\rho)=\left[
\begin{array}{cc}
\cos(u_{j} \Delta x_{j}) & -\sin(u_{j} \Delta x_{j})\,\sigma_{j}/u_{j} \\
&  \\ 
\sin(u_{j} \Delta x_{j})\; u_{j}/\sigma_{j} &
\cos(u_{j} \Delta x_{j})  \label{cs}
\end{array}
\right]\;\;.
\end{equation}
${\bf X}(x)$ and $\tilde{\bf X}(x)$ represent vectors containing the modal functions $X_{m}(x)$ and
$\tilde{X}_{m}(x)$, respectively, and $u$ and $\rho$ are related by the dispersion relation. The subscript $m$ denoting the mode has
been suppressed as it is understood that relations (\ref{uu})-(\ref{cs}) hold for each one of
the modal terms.  This procedure can be applied N times to get a relation
between the fields at the perfectly conducting walls at $x=x_{0}$ and
$x=x_{L}$. In such a case we have a relation of the form:
\begin{equation}
\left(
\begin{array}{cc}
{\bf X}(x_{L})\\
\\ 
\tilde{\bf X}(x_{L})
\end{array}
\right) =
\tilde{\bf M}(\rho)
\left(
\begin{array}{cc}
{\bf X}(x_{0}) \\
\\ 
\tilde{\bf X}(x_{0})
\end{array}
\right)\;\;,  \label{relmfin}
\end{equation}
where $\tilde{\bf M}(\rho)$ is a product matrix:
\begin{equation}
\tilde{\bf M}(\rho)={\bf M}_{L}(\rho)\;
{\bf M}_{L-1}(\rho)\;\ldots\;
{\bf M}_{2}(\rho)\;
{\bf M}_{1}(\rho)\;\;.  \label{matm}
\end{equation}
$\tilde{\bf M}(\rho)$ is a $2 \times 2$ block matrix. This is a
well known result from the theory of stratified media \cite{Brekhovskikh}.

Imposition of the boundary condition at $x_{0}$ and at $x_{L}$, i.e.,
the tangential component of the electric field must vanish on the surface,
yields a condition on the non-diagonal block elements: 
\begin{eqnarray}
\tilde{\bf M}_{12}(\rho)&=&0 \;\;\;\;\;\mbox{for TM polarization} \label{mtm}\\
\tilde{\bf M}_{21}(\rho)&=&0 \;\;\;\;\;\mbox{for TE polarization} \label{mte}\;\;.
\end{eqnarray}
These conditions determine eigenvalues
equations for $\rho$, that must be solved by means of numerical techniques. 
These equations have already been studied for structures formed by isotropic regions 
in the case of an infinite periodic grating \cite{Miller} and of an index-modulated aperture 
\cite{Kuittinen, diana10}. The expressions of the equations found in this work reduce
to those in Refs. \cite{Kuittinen, diana10} for $b=1$. Also, a recursive formula to find
the eigenvalues equation of a layer with N different regions could be derived in
the present case \cite{diana10}.
Finally, the eigenmodes of the layer
bounded by perfectly matched layers backed by perfectly conducting walls are given by
\begin{eqnarray}
X_m(x) = A_{mj} \cos[u_{mj} (x-x_j)] + B_{mj} \sin[u_{mj} (x-x_j)] 
\;\;\;\mbox{for $x_j\,<\,x\,<\,x_{j}+\Delta_j$}\;\;,      \label{xmx}
\end{eqnarray}
where $u_{mj}$ satisfy the corresponding eigenvalues equation (\ref{mtm}) or (\ref{mte}).
The modal coefficients $A_{mj}$ and $B_{mj}$ are easily obtained by successive
application of eqs. (\ref{uu})
\begin{eqnarray}
A_{mj}&=&A_{m,j-1} \cos[u_{m,j-1} (x_j-x_{j-1})] + B_{m,j-1} \sin[u_{m,j-1} (x_j-x_{j-1})] \nonumber \\
B_{mj}&=&\frac{u_{m,j-1}}{u_{mj}} \frac{\sigma_j}{\sigma_{j-1}} \left\{B_{m,j-1} \cos[u_{m,j-1} (x_j-x_{j-1})]
-A_{m,j-1} \sin[u_{m,j-1} (x_j-x_{j-1})]\right\}\;\;.   \label{amybm}
\end{eqnarray}

\section{Examples}

The eigenvalues equation is the critical part of any modal formulation, particularly
for layers comprising different kinds of materials.
Equations of this kind have been solved in Refs. \cite{Miller}, \cite{Kuittinen} and
\cite{diana9} for real refractive indices (non lossy dielectrics) and in \cite{diana12}
for lossy materials. A detailed study of the respective equations has been performed
in the above works to avoid numerical problems with the rootfinding algorithms. In
\cite{Kuittinen},\cite{diana9} and \cite{diana10} the authors use the separation 
constant along 
the $y$ direction ($v^2$) as the variable of the equation, whereas for highly conducting
metals, i.e. regions of refraction index with a large imaginary part, it is 
convenient to use the separation constant in the $x$ direction ($u$) as the variable 
\cite{diana12}. This behaviour can be understood taking into account the location of the eigenvalues in two limit cases: (i) ideal dielectric isotropic material (real refraction index), and (ii) perfect conductor (infinite refraction index). In the first case, the 
eigenvalues ($v^2$) are real, and then, for small perturbations of the refraction index (lossy dielectrics) a small departure from these values is expected. 
On the contrary, for the perfect conductor the eigenvalues ($u$) are real and can be calculated analytically \cite{diana2}. Therefore, we expect the
eigenvalues corresponding to a highly conducting metal not to depart too much from
those real values. 

As a first example, we consider the case of a symmetric structure formed by three zones (PML - dielectric - PML) of widths $\Delta_{P}$ and $\Delta_1$, and then $u_0=u_2=u_{P}$. This example actually simulates the open space, and can be useful to study the coupling between a slab waveguide and open space 
\cite{Derudder2}. For this case, the direct application of the procedure described in the previous 
section yields an eigenvalues equation for both polarization modes, that after some manipulation can 
be reduced to:
\begin{equation}
\sin(2 u_{P} \Delta_{P} + u_1 \Delta_1)=0\;\;.   \label{autov-caso1}
\end{equation}
Taking into account that $u_{P}=u_1 b$, the solutions of eq. (\ref{autov-caso1}) are
\begin{equation}
u_{1m}=\frac{m \pi}{2\, b\, \Delta_{P}+\Delta_1}\;\;,\;\;\;\mbox{for $m$ integer}.		\label{u1-caso1}
\end{equation}
The eigenvalues obtained in (\ref{u1-caso1}) are the same as those found by Derudder {\em et al.} by 
means of a coordinates transformation in the propagation equation (eqs. (16) and (17) in
Ref. \cite{Derudder2}).

\begin{figure}[h]
\begin{center}
\includegraphics[width=5in]{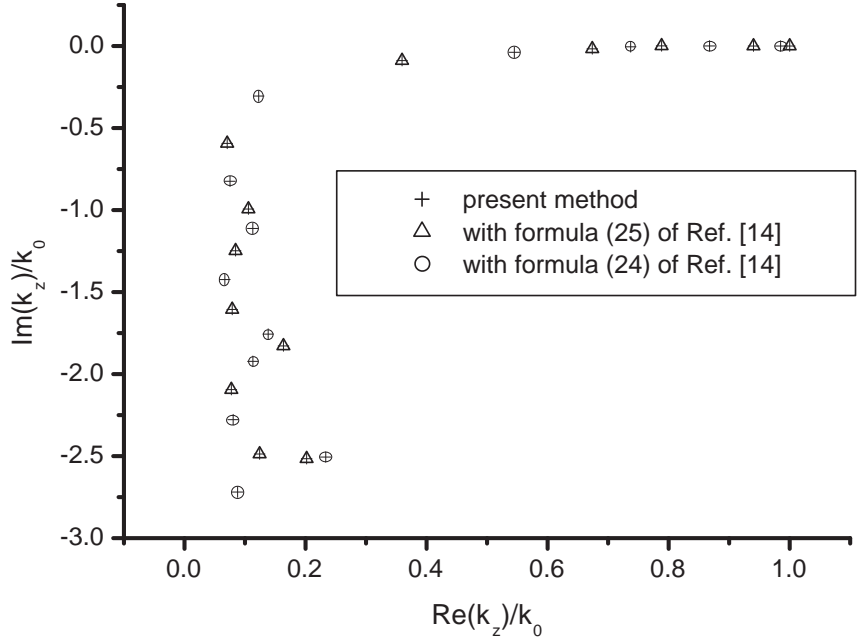}
\end{center}
\caption{Comparison of the eigenvalues obtained by the present method and that of Derudder {\em et. al}
(eqs. (24) and (25) in Ref. \cite{Derudder2}) for a structure with 5 zones: $\Delta_P/\lambda=0.04$, 
$\Delta_1/\lambda=\Delta_3/\lambda=0.5$, $\Delta_2/\lambda=2$, $n_1=n_3=1$, $n_2=1.3$, $b=2-2i$, TM polarization.}
\end{figure}

It is important to remark that an increase in the number of homogeneous regions in the 
layer produces a more complex eigenvalues equation, since it is obtained from the determinant 
of a matrix 
product (eq. (\ref{matm})). The number of terms and the number of trigonometric
factors in each term increases, causing abrupt oscillations in the equation. However,
since we are focusing on the solution of scattering problems from homogeneous objects
in open space, the number of different regions to be considered in this case, including the PML, 
is restricted to five. The explicit expressions of the eigenvalues equations for five zones 
and for $\Delta_1=\Delta_3$ and $\mu_j=\mu_0 \forall j$ are
given in the Appendix for TE and TM polarizations. As it can be observed 
in Fig. 3 for TM polarization, the solutions of the equation are coincident with 
those obtained from eqs. (24) and (25) in Ref. \cite{Derudder2}. These equations correspond
to the eigenvalues equations for odd and even modes, respectively. 

In Fig. 4 we show the evolution of the eigenvalues as the imaginary part of the PML parameter
$b$ is increased, for a structure comprising five zones and TE polarization. For $Im(b)=0$,
the eigenvalues $u_1$ are real, and their values depart from the real axis as $Im(b)$ is increased. 

\begin{figure}[h]
\begin{center}
\includegraphics[width=5in]{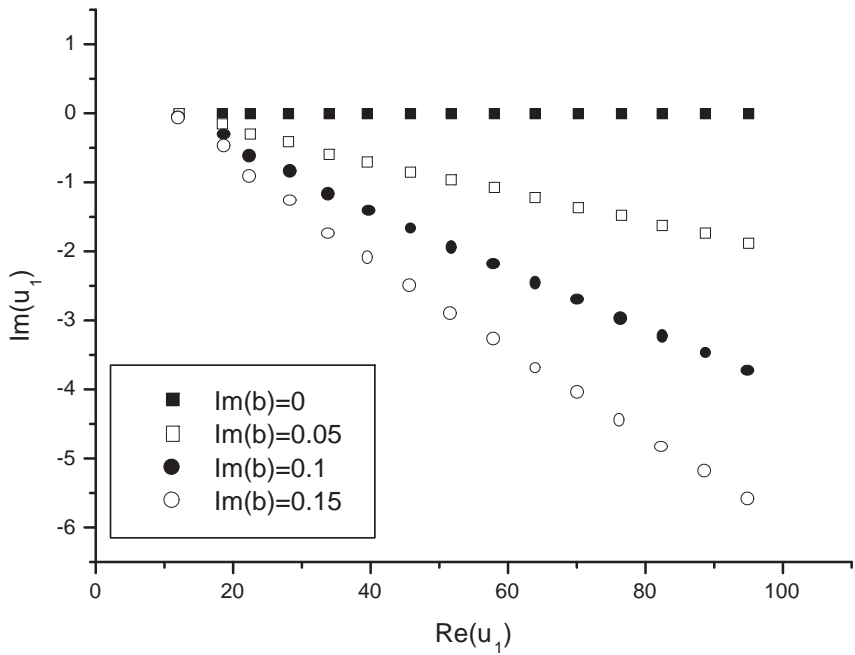}
\end{center}
\caption{Evolution of the eigenvalues with the imaginary part of the PML parameter $b$, for a 
structure with 5 zones: $\Delta_P/\lambda=\Delta_1/\lambda=\Delta_2/\lambda=\Delta_3/\lambda=0.2$, 
$n_1=n_3=1$, $n_2=1.5$, $Re(b)=1$ and TE polarization.}
\end{figure}

As it has been stated above, any system of planar regions can be studied by the modal method.
If we take into account that any profile can be approximated by a stack of
rectangular layers (multilayer approximation) \cite{Peng2}, the scope of applicability of
the modal method broadens significantly. Since we have already found the exact expression of 
the field in terms of the eigenmodes of a rectangular layer, the method can be applied to a
multilayer system by means of any propagation algorithm such as the T-matrix, R-matrix, 
and the S-matrix \cite{Li2}.
The procedure is essentially that formulated in Refs. \cite{diana9},\cite{diana10} for 
dielectric apertures. The main difference is that in the present case, i.e., when the 
layers are ended by PMLs, the
definition of the modal functions $U_{m,j}^q (x)$ (eq. (9) in Ref. \cite{diana10}) should be
substituted by the expression of $X_m(x)$ given in eq. (\ref{xmx}) for 
each layer.

\section{Summary and conclusions}

Analytical expressions for the eigenmodes of an inhomogeneous layer ended by particular 
aniso\-tro\-pic regions (PMLs) have been found. The general method was described
for an arbitrary number of regions of different widths and dielectric materials in the structure. 
The eigenvalues equation for structures comprising lateral anisotropic lossy 
regions was found, what opens up the possibility of application of the modal method to simulate 
scattering problems from 2-D objects in open space. A complete set of
modal functions was obtained, and also the orthonormality relation between them was derived. 
Simple examples of eigenvalues equations for planar slab waveguides (three and five zones
in the layer) have been presented, and the results
have been compared with those found in the literature. The present formulation constitutes an
extension of the modal method and it is a 
basic tool that can be applied to more general problems such as propagation in waveguides
of arbitrary cross section and scattering problems from arbitrarily shaped objects.

\section*{Acknowledgments}
D. S. acknowledges Dr. Miriam Gigli for the rootfinding routine.

The author gratefully acknowledges partial support from Consejo Nacional de
Investigaciones Cient\'{\i}ficas y T\'ecnicas (CONICET) and Universidad de Buenos Aires (UBA). 

\section*{Appendix}
Explicit expressions for the eigenvalues equations obtained for a symmetric structure of five zones
(PML - 1 - 2 - 1 - PML), with $\mu_j=\mu_0 \forall j$.

\underline{TM polarization}

\begin{eqnarray}
&&2 \cos(u_{P} \Delta_{P}) \cos(u_1 \Delta_1) \sin(u_{P} \Delta_{P}) [2 \sin(u_2 \Delta_2)^2
-1] + \nonumber \\
&&2 \sin(u_2 \Delta_2) \cos(u_2 \Delta_2) \cos(u_1 \Delta_1) [2 \sin(u_{P} \Delta_{P})^2-1] + \nonumber \\
&&2 \cos(u_{P} \Delta_{P})\sin(u_2 \Delta_2)\cos(u_2 \Delta_2)\sin(u_1 \Delta_1)
\sin(u_{P} \Delta_{P})\left(\frac{u_1}{u_2}+\frac{u_2}{u_1}\right)+ \nonumber \\
&&\sin(u_{P} \Delta_{P})^2
\sin(u_1 \Delta_1)\left[-\sin(u_2 \Delta_2)^2\frac{u_2}{u_1}+\cos(u_2 \Delta_2)^2\frac{u_1}{u_2}\right]+
\nonumber \\
&&\cos(u_{P} \Delta_{P})^2\sin(u_1 \Delta_1)\left[-\cos(u_2 \Delta_2)^2\frac{u_2}{u_1}+
\sin(u_2 \Delta_2)^2\frac{u_1}{u_2}\right]=0 
\end{eqnarray}

\underline{TE polarization}
\begin{eqnarray}
&&2 \sin(u_{P} \Delta_{P}) \cos(u_{P} \Delta_{P}) \cos(u_{1} \Delta_{1}) [1-2 \sin(u_2 \Delta_2)^2] + 
\nonumber \\
&&2 \sin(u_2 \Delta_2) \cos(u_2 \Delta_2) \cos(u_1 \Delta_1) [1-2 \sin(u_{P} \Delta_{P})^2] - \nonumber \\
&&2 \cos(u_{P} \Delta_{P})\sin(u_2 \Delta_2)\cos(u_2 \Delta_2)\sin(u_1 \Delta_1)
\sin(u_{P} \Delta_{P})\left(\frac{u_1 \,\epsilon_2}{u_2 \,\epsilon_1}+ 
\frac{u_2 \,\epsilon_1}{u_1\, \epsilon_2}\right)\nonumber -\\
&&\frac{u_2 \,\epsilon_1}{u_1\, \epsilon_2}\sin(u_1 \Delta_1)
\left[\sin(u_{P} \Delta_{P})^2\cos(u_2 \Delta_2)^2+\sin(u_2 \Delta_2)^2 \cos(u_{P} \Delta_{P})^2\right]+\nonumber \\
&&\frac{u_1\, \epsilon_2}{u_2\, \epsilon_1}
\sin(u_1 \Delta_1)\left[\cos(u_2 \Delta_2)^2 \cos(u_{P} \Delta_{P})^2+
\sin(u_2 \Delta_2)^2 \sin(u_{P} \Delta_{P})^2 \right]=0
\end{eqnarray}


\newpage
\begin{thebibliography}{99}

\bibitem{berenger1} J. P. Berenger, ``A perfectly matched layer for the
absorption of electromagnetic waves'' J. Comput. Phys.  114,  185-200 (1994).

\bibitem{berenger2} J. P. Berenger, ``Three-Dimensional Perfectly Matched
Layer for the Absorption of Electromagnetic Waves'', J. Comput. Phys. 127, 363-379 (1996).

\bibitem{Sacks} Z. S. Sacks, D. M. Kingsland, R. Lee and J. F. Lee, ``A
perfectly matched anisotropic absorber for use as an absorbing boundary
condition'', IEEE Trans. Antennas Propag. 43 1460-1463 (1995).

\bibitem{PML_JOSAB} Jeong-Ki Hwang, Seok-Bong Hyun, Han-Youl Ryu, 
Yong-Hee Lee, ``Resonant modes of two-dimensional photonic bandgap cavities
determined by the finite-element method and by use of the anisotropic
perfectly matched layer boundary condition'', J. Opt. Soc. Am. B 15 2316-2324 (1998) .

\bibitem{PML_AO} Wenbo Sun, Qiang Fu, Zhizhang Chen, ``Finite-difference 
time-domain solution of light scattering by dielectric particles with a perfectly 
matched layer absorbing boundary condition'', Appl. Opt. 38 3141-3151 (1999).
 
\bibitem{Merle1} J. Merle Elson, P. Tran, ``R-matrix propagator with 
perfectly matched layers for the study of integrated optical components'', J. Opt. 
Soc. Am. A 16 2983-2989 (1999).

\bibitem{Merle2} J. Merle Elson, ``Propagation in planar waveguides and the
effects of wall roughness'', Opt. Express 9, 461-475 (2001).

\bibitem{PML2} D. C. Skigin and R. A. Depine, ``Use of an anisotropic absorber for 
simulating electromagnetic scattering by a perfectly conducting wire'', Optik 114 (5), 
229-233 (2003).

\bibitem{Rogier1} H. Rogier, D. De Zutter, ``Convergence behavior and acceleration
of the Berenger and leaky modes series composing the 2-D Green's function for the microstrip substrate'', IEEE Trans. Microwave Theory and Tech. 50, 1696-1704 (2002).

\bibitem{Rogier2} H. Rogier, D. De Zutter, ``A fast technique based on perfectly 
matched layers for the full-wave solution of 2-D dispersive microstrip lines'', IEEE Trans. Computer Aided Design of Integrated Circuits and Systems 22, 1650-1656 (2003).

\bibitem{Rogier3} H. Rogier, D. De Zutter, ``A fast converging series expansion for the 2-D periodic Green's function based on perfectly matched layers'', IEEE Trans. Microwave Theory and Tech. 52, 1199-1206 (2004).

\bibitem{Gedney} S. D. Gedney, ``An anisotropic perfectly matched layer-absorbing
medium for the truncation of FDTD lattices'', IEEE Trans. Antennas Propag. 44, 1630-1639 (1996).

\bibitem{Mekis} A. Mekis, S. Fan and J. D. Joannopoulos, ``Absorbing boundary conditions
for FDTD simulations of photonic crystal waveguides'', IEEE Microwave Guided Wave Lett. 9,
502-504 (1999).

\bibitem{Tischler} T. Tischler and W. Heinrich, ``The perfectly matched layer as lateral
boundary in finite-difference transmission-line analysis'', IEEE Trans. Microwave Theory 
Tech. 48, 2249-2253 (2000).

\bibitem{Derudder0} H. Derudder, D. De Zutter and F. Olyslager, ``Analysis of waveguide
discontinuities using perfectly matched layers'', Electron. Lett. 34, 2138-2140 (1998). 

\bibitem{Derudder1} H. Derudder, F. Olyslager and D. De Zutter, ``An efficient series 
expansion for the 2D Green's function of a microstrip substrate using perfectly matched
layers'', IEEE Microwave Guided Wave Lett. 9, 505-507 (1999).

\bibitem{Derudder2} H. Derudder, F. Olyslager, D. De Zutter and S. Van den Berghe,
``Efficient mode-matching analysis of discontinuities in finite planar substrates
using perfectly matched layers'', IEEE Trans. Antennas Propag. 49, 185-195 (2001).

\bibitem{Rogier4} H. Rogier, D. De Zutter, ``Berenger and leaky modes in microstrip 
substrates terminated by a perfectly matched layer'', IEEE Trans. Microwave 
Theory Tech. 49, 712-715 (2001).

\bibitem{Olyslager1} F. Olyslager, H. Derudder, ``Series representation of Green dyadics 
for layered media using PMLs'', IEEE Trans. Antennas Propag. 51, 2319-2326 (2003).

\bibitem{Knop} K. Knop, ``Rigorous diffraction theory for transmission phase gratings
with deep rectangular grooves'', J. Opt. Soc. Am. 68, 1206 (1978).

\bibitem{Peng2}  S. T. Peng, T. Tamir and H. L. Bertoni, ``Theory of periodic
dielectric waveguides'', IEEE Trans. Microwave Theory Tech. MTT-23, 123-133
(1975).

\bibitem{Botten0} L. C. Botten, M. S. Craig, R. C. McPhedran, J. L. Adams
and J. R. Andrewartha, ``The dielectric lamellar diffraction grating'', Opt. Acta 28, 413-428 (1981).

\bibitem{Peng1} S. T. Peng, ``Rigorous formulation of scattering and guidance by dielectric grating waveguides: general case of oblique incidence'', J. Opt. Soc. Am. A6, 1869-1883 (1989). 

\bibitem{Li} L. Li, ``A modal analysis of lamellar diffraction
gratings in conical mountings'', J. Mod. Optics 40, 553-573 (1993).

\bibitem{Miller} J. M. Miller, J. Turunen, E. Noponen, A. Vasara and M. R.
Taghizadeh, ``Rigorous modal theory for multiply grooved lamellar gratings'',
Opt. Commun. 111, 526-535 (1994).

\bibitem{Kuittinen} M. Kuittinen and J. Turunen, ``Exact-eigenmode model for index-modulated
apertures'', J. Opt. Soc. Am. A13, 2014-2020 (1996).

\bibitem{Li2} L. Li, ``Multilayer modal method for diffraction gratings of
arbitrary profile, depth and permittivity'', J. Opt. Soc. Am. A10, 2581-2591
(1993).

\bibitem{diana9} R. A. Depine and D. C. Skigin, ``Multilayer modal method for 
diffraction from dielectric inhomogeneous apertures'', J. Opt. Soc. Am. A 15, 675-683 (1998). 

\bibitem{diana10} D. C. Skigin and R. A. Depine, ``Modal theory for diffraction
from a dielectric aperture with arbitrarily shaped corrugations'', Opt.
Commun. 149, 1-3, 117-126 (1998).

\bibitem{diana12} D. C. Skigin and R. A. Depine, ``Scattering by lossy inhomogeneous 
apertures in thick metallic screens'', J. Opt. Soc. Am. A 15, 2089-2096 (1998).

\bibitem{Olyslager2} F. Olyslager, ``Discretization of continuous spectra based
on perfectly matched layers, SIAM J. Appl. Math. 64, 1408-1433 (2004).

\bibitem{Knockaert} L. F. Knockaert, D. De Zutter, ``On the completiness of
eigenmodes in a parallel plate waveguide with a perfectly matched layer termination'',
IEEE Trans. Antennas Propag. 50, 1650-1653 (2002).

\bibitem{Li3} L. Li, ``Multilayer-coated diffraction gratings:
differential method of Chandezon {\em et al.} revisited'', J. Opt. Soc. Am. A
11, 2816-2828 (1994).

\bibitem{Li4} L. Li, ``Bremmer series, R-matrix propagation algorithm, and
numerical modelling of diffraction gratings'', J. Opt. Soc. Am. A 11,
2829-2836 (1994).

\bibitem{Neviere} F. Montiel and M. Nevi\`ere, ``Differential theory of gratings:
extension to deep gratings of arbitrary profile and permittivity through the
R-matrix propagation algorithm'', J. Opt. Soc. Am. A 11, 3241-3250
(1994).

\bibitem{Jackson} J. D. Jackson, ``Classical Electrodynamics'', 2nd. ed.,
Wiley, New York (1975).

\bibitem{DiffEcs} R. H. Cole, ``Theory of ordinary differential equations'', Appleton-Century-Crofts,
New York (1968).

\bibitem{Brekhovskikh} L. M. Brekhovskikh, {\it Waves in Layered Media},
Academic, New York, 1960.

\bibitem{diana2} R. A. Depine and D. C. Skigin, ``Scattering from metallic
surfaces having a finite number of rectangular grooves'', J. Opt. Soc. Am.
A11, 2844-2850 (1994). 

\end{thebibliography}
\end{document}